\documentclass[%
 aip,
 amsmath,amssymb,
 reprint,%
]{revtex4-1}

\usepackage{graphicx}
\usepackage{dcolumn}
\usepackage{bm}
\usepackage[utf8]{inputenc}
\usepackage[T1]{fontenc}
\usepackage{mathptmx}
\usepackage{etoolbox}
\usepackage{changes}
\definechangesauthor[color=blue]{Rev}

\makeatletter
\def\@email#1#2{%
 \endgroup
 \patchcmd{\titleblock@produce}
  {\frontmatter@RRAPformat}
  {\frontmatter@RRAPformat{\produce@RRAP{*#1\href{mailto:#2}{#2}}}\frontmatter@RRAPformat}
  {}{}
}%
\makeatother
\begin{document}

\preprint{AIP/123-QED}

\title[Dynamics in Water-Ethanol Mixtures]{Molecular Structural Dynamics in Water-Ethanol Mixtures:\\ Spectroscopy with Polarized Neutrons Simultaneously Accessing Collective and Self-Diffusion}
\author{Riccardo Morbidini}
\affiliation{Institut Max von Laue - Paul Langevin, 71 Avenue des Martyrs, F-38042 Grenoble, France.}
\affiliation{Division of Pharmacy and Optometry, University of Manchester, Oxford Road, Manchester M13 9PT, United Kingdom.}
\author{Robert M.~Edkins}%
\affiliation{WestCHEM Department of Pure and Applied Chemistry, University of Strathclyde, 295 Cathedral Street, Glasgow G1 1XL, United Kingdom.}
\author{Mark Devonport}
\affiliation{ISIS Neutron and Muon Source, Rutherford Appleton Laboratory, Didcot, OX11 0QX, United Kingdom.}
\author{G{\o}ran Nilsen}
\affiliation{ISIS Neutron and Muon Source, Rutherford Appleton Laboratory, Didcot, OX11 0QX, United Kingdom.}
\author{Tilo Seydel*}
\affiliation{Institut Max von Laue - Paul Langevin, 71 Avenue des Martyrs, F-38042 Grenoble, France.}

\author{Katharina Edkins*}
\email[Corresponding authors' electronic mail: ]{seydel@ill.eu, katharina.edkins@manchester.ac.uk.}
\affiliation{Division of Pharmacy and Optometry, University of Manchester, Oxford Road, Manchester M13 9PT, United Kingdom.}%


\begin{abstract}
Binary mixtures of water with lower alcohols display non-linear phase behaviour upon mixing which are attributed to potential cluster formation at molecular level. Unravelling such elusive structures requires the investigation of hydrogen-bonding sub-nanosecond dynamics. We employ high-resolution neutron time-of-flight spectroscopy with polarization analysis in combination with selective deuteration to study the concentration-dependent structural dynamics, in the water rich part of the phase diagram of water-ethanol mixtures. This method enables the simultaneous access to atomic correlations in space and time, and allows us to separate spatially incoherent scattering probing self-diffusion of the ethanol fraction from the coherent scattering probing collective diffusion of the water network as a whole. Our observations indicate an enhanced rigidity of the hydrogen bond network at mesoscopic lengthscale compared to the intra-molecular scale as the ethanol fraction increases, which is consistent with the hypothesis of clusters.
\end{abstract}

\maketitle
Due to their role in food and pharmaceutical industry, as well as for biological and chemical research, water~\cite{ball2008water} and its binary mixtures with monohydroxy alcohols~\cite{franks1966structural} have been the subject of sustained focus. Water and lower alcohols, such as methanol, ethanol and 1-propanol,~\cite{grossmann1981formation} 
display intriguing properties e.g. negative excess entropy,~\cite{soper2006excess} non-ideal phase behaviour upon mixing, and non-monotonous dependencies of volume, refractive index and viscosity on the mixing ratio.~\cite{gonzalez2007density,herraez2006refractive}
These macroscopic properties relate to the molecular complexity of the water hydrogen-bond network perturbed by the amphiphilic nature of alcohol molecules.~\cite{guo2003molecular,noskov2005molecular} 
The microheterogeneity at the molecular level is commonly interpreted in terms of local clusters,~\cite{povzar2016micro,dixit2002molecular,wakisaka2006microheterogeneity, gereben2017cluster} whose topology and temperature, pressure and concentration dependence has been systematically examined through structural methods such as X-ray and neutron diffraction,~\cite{lenton2018temperature,pethes2020temperature,dougan2004methanol,takamuku2005x} nuclear magnetic resonance spectroscopy (NMR)~\cite{price1999self} or simulations.~\cite{gereben2017cluster,pothoczki2018temperature} However, the structural description is not accompanied by a comprehensive dynamic profile,~\cite{lovrincevic2021overview} due to the difficulty of tracking the elusive nature of H-bonded associates, whose lifetimes are typically on the order of picoseconds or less.~\cite{jukic2022lifetime}
Pulsed-field gradient NMR spectroscopy (PFG-NMR) allows the measurement of the molecular self-diffusion~\cite{price1999self} despite averaging over the cluster fluctuations, which are $\sim$10$^{9}$ times faster than the NMR time resolution (millisecond). MD simulations match the desired observation scale for self diffusion  
of the individual species although the choice of the intermolecular force fields need experimental support.~\cite{obeidat2018validity,obeidat2018validity_eth,perez2021molecular,bako2008water,wensink2003dynamic} Along with the self-diffusion of individual species, studying the dynamics of the H-bonded network as a collective ensemble is equally important. Dielectric spectroscopy (DS) has been widely used to identify collective dipole relaxation of water-ethanol mixtures.~\cite{bohmer2014structure,wieth2014dynamical,sato2004dielectric,gainaru2010nuclear,li2014mesoscopic,sillren2014liquid}
Here, a low-frequency mode (corresponding to a timescale of $\sim$10 ps) has been attributed to the cooperative dynamics of the whole network which becomes less flexible with increasing ethanol ratio.~\cite{kaatze2002hydrogen,cardona2018molecular} However, these observations do not contain direct information on the spatial correlations.
Neutron spectroscopy simultaneously accesses spatial and time correlations at the molecular level by measuring momentum transfer $\hbar q$ and energy transfer $\hbar\omega$ and provides an ideal probe complementing NMR, MD simulations and DS. 
\\
Quasi-elastic neutron scattering (QENS) has been successfully employed to study the single-particle self-dynamics in pure water~\cite{teixeira1985experimental,qvist2011structural} and in water-ethanol mixtures,~\cite{seydel2019picosecond} focusing on the ensemble-average response of hydrogen atoms that dominate the incoherent scattering.
In contrast, measurements of collective motions have been limited to the ns timescale~\cite{luo2022q} with their impact on the ps timescale often minimized by limiting the coherent scattering length in the samples. 
Polarization analysis (PA)~\cite{chen2019wideangle,nilsen2017polarisation} is the only way to unambiguously separate the coherent and incoherent QENS signals and it has already been applied to H-bonded liquids~\cite{burankova2014collective,bertrand2016dynamic,bertrand2017nanoscopic} building on theoretical concepts of collective density fluctuation in liquids on the mesoscale ($0.4\,$\AA$^{-1}$$\leq q\leq 1.9\,$\AA$^{-1}$).~\cite{novikov2013coherent}
The recent progress of polarization analysis on cold neutron spectrometers with sub-meV resolution allowed the heterogeneous coherent structural relaxation of water (D$_2$O) to be unravelled.~\cite{arbe2020coherent,arbe2022understanding}
Building on this result, we combine selective deuteration and PA to simultaneously measure self-diffusion of individual species in water-ethanol mixtures while analysing collective density fluctuations as a function of alcohol concentration.
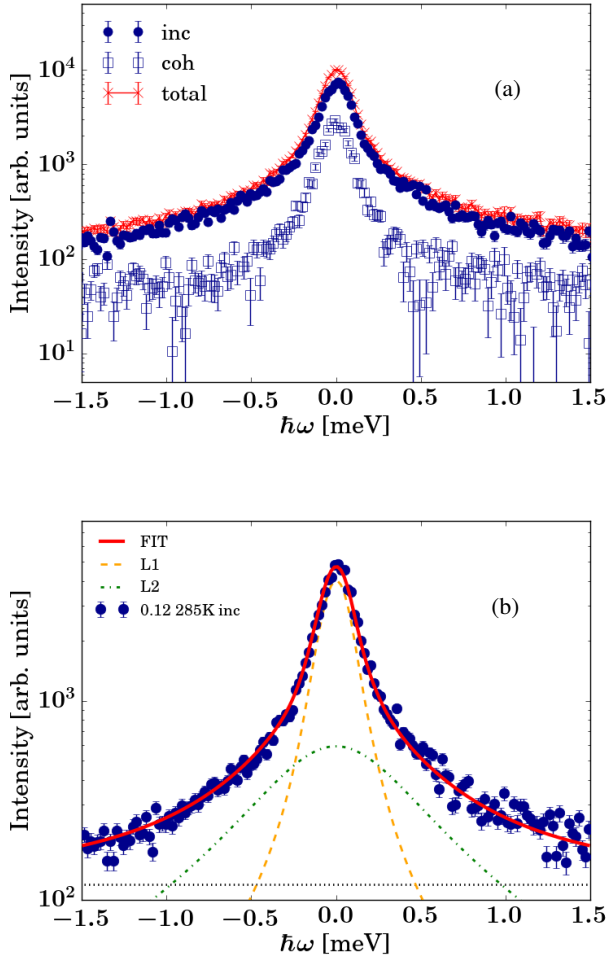
\begin{figure}[t!]
    \centering
    \centering
    {\includegraphics[width=\columnwidth]{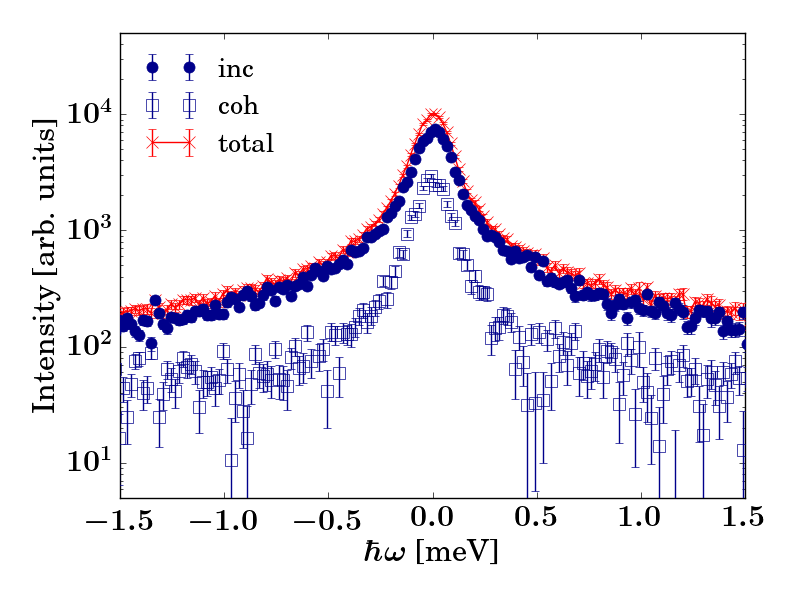}}
    \begin{picture}(0,0)
    \put(70,150){(a)}        
    \end{picture}
    \label{cohinctot}
    \hfill    
    \centering
    {\includegraphics[width=\columnwidth]{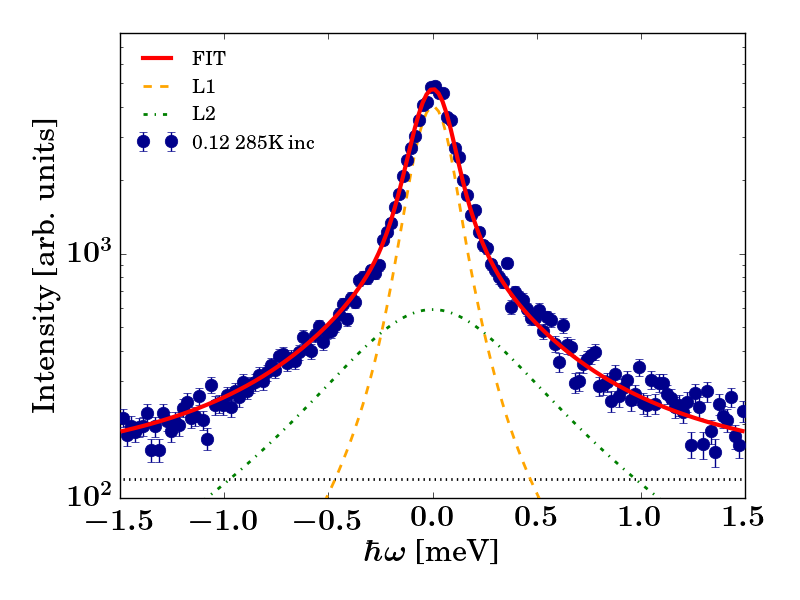}}
    \begin{picture}(0,0)
    \put(70,150){(b)}        
    \end{picture}
    \label{fitexample}
    \caption{\footnotesize (a) Comparison of the coherent (open squares), incoherent (full circles) and total (red) scattering functions measured on LET on D$_{2}$O/C$_{2}$H$_{5}$OD at 0.12 ethanol mole fraction, T=285 K at $q$=1.6\,\AA$^{-1}$. (b) Example incoherent scattering function at $q$=1.4\,\AA$^{-1}$ recorded on LET at E$_{\mathrm{i}}$=3.84 meV. Fit (red solid line) according to eq.\ref{fit model} consisting of a sum of two Lorentzians (dashed and dash-dotted line) and an almost negligible flat apparent background (dotted line). }
    \label{fig_expl_spec}
\end{figure}
\begin{figure}[h!]
\centering
\includegraphics[width=1\columnwidth]{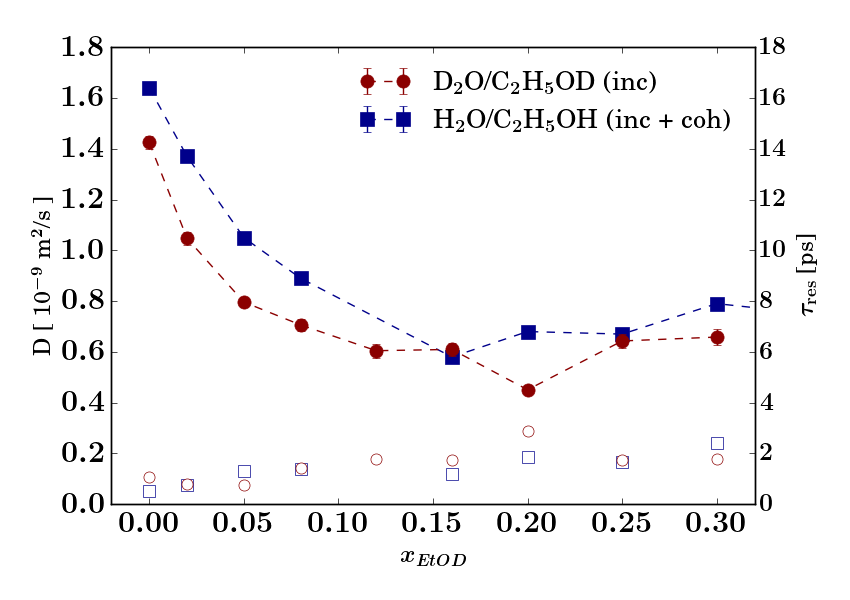}
\caption{\footnotesize\label{fig_DandTau} Diffusion coefficients $D$ (full red circles) and residence times $\tau_\mathrm{res}$ (open red circles) of D$_2$O/C$_{2}$H$_{5}$OD mixtures as a function of ethanol mole fraction $x_{EtOD}$ at T=285 K and E$_{\mathrm{i}}$=3.84 meV retrieved by imposing a jump diffusion model eq.\ref{jump} on the HWHM vs $q^{2}$ of the fitted $S_{\mathrm{inc}}(q,\omega)$. Hydrogen-rich mixtures H$_2$O/C$_{2}$H$_{5}$OH previously studied on LET without polarization analysis is reported for comparison (full blue squares $D$, open blue squares $\tau_\mathrm{res}$).~\cite{seydel2019picosecond} 
} 
\end{figure}
%
We performed neutron spectroscopy with polarization analysis on the cold neutron time-of-flight spectrometer LET at the ISIS facility, Rutherford Appleton Laboratory, Didcot, UK.~\cite{nilsen2017polarisation,bewley2011let} The multi-chopper system of this direct geometry instrument provides three different sequentially incident energies E$_{i}$ of 3.84, 1.81 and 1.05 meV in a single measurement associated with three distinct energy resolutions, respectively $\Delta$E$\approx131\mu$eV (3.4$\%$E$_{i}$), $\Delta$E$\approx45\mu$eV (2.5$\%$E$_{i}$) and $\Delta$E$\approx22\mu$eV (2.1$\%$E$_{i}$) FWHM (example plots in the SI).
We obtain separate QENS spectra for the coherent $S_{\mathrm{coh}}(q,\omega)$ and incoherent $S_{\mathrm{inc}}(q,\omega)$ scattering functions. We used specific deuteration, mixing deuterium oxide (D$_2$O) and ethanol-OD (C$_2$H$_5$OD) to distinguish water and ethanol at different ethanol mole fractions (0.0, 0.02, 0.05, 0.08, 0.12, 0.16, 0.20, 0.25, 0.30) near the maximum of macroscopic viscosity.
Therefore, $S_{\mathrm{coh}}(q,\omega)$ will represent D$_2$O dynamics, while the five non-exchangeable hydrogens of C$_{2}$H$_{5}$OD give the higher incoherent scattering cross section and dominate $S_{\mathrm{inc}}(q,\omega)$. In addition, a mixture of D$_2$O and ethanol C$_2$D$_5$OD was measured to account for the effect of selective deuteration on the coherent structural relaxation of the hydrogen bond (HB) network. 
Examples of coherent and incoherent QENS spectra are depicted in Fig.~\ref{fig_expl_spec}(a) displaying the comparison of the collected coherent, incoherent and calculated total scattering. While for pure D$_2$O the coherent term is dominant over the incoherent especially at $q$ near the peak of the structure factor $S(q)$,~\cite{arbe2020coherent} 
already at low alcohol concentration the coherent term is masked throughout the whole $q$ range by the incoherent response arising from the non-exchangeable hydrogens of ethanol (see Fig.~\ref{fig_expl_spec}(a) and Fig.S1). Thus, only polarization analysis allows $S_{\mathrm{coh}}(q,\omega)$ to be accessed and at the same time to circumvent the De Gennes narrowing~\cite{degennes1959liquid} which may affect the $S_{\mathrm{inc}}(q,\omega)$ when not separated from the $S_{\mathrm{coh}}(q,\omega)$. It has been shown how this leads to an underestimation of the self diffusion coefficient when there is no clear separation between coherent and incoherent spectra.~\cite{burankova2014collective} \\
Firstly we focused on the self-diffusion by fitting the reduced incoherent spectra with $S_{\mathrm{inc}}(q,\omega)\otimes\mathcal{R}(q,\omega)$ 
wherein the spectrometer resolution function $\mathcal{R}(q,\omega)$ is obtained by fitting a vanadium sample (Fig.~S2) for each $q$ independently and $S_{\mathrm{inc}}(q,\omega)$ is described by well established model for H$_2$O,~\cite{qvist2011structural}
\begin{eqnarray}\nonumber
    S_{\rm inc}(q,\omega)=&a_{1}(q)\frac{1}{\pi}\frac{\displaystyle \gamma_{1}(q)}{\displaystyle \gamma_{1}(q)^{2}+\omega^{2}}&\\
    &+ a_{2}(q)\frac{1}{\pi}\frac{\displaystyle \gamma_{2}(q)}{\displaystyle \gamma_{2}(q)^{2}+\omega^{2}}&+b(q),
    \label{fit model}
\end{eqnarray}
consisting of a sum of two Lorentzian functions with amplitudes $a_{i}(q)$ and half width at half maximum (HWHM) $\gamma_{i}(q)$ accounting for slow translational process (\textit{i}=1) and fast motions (\textit{i}=2) with a small flat background $b(q)$ arising from instrument, sample, and sample environment contributions (Fig.\ref{fig_expl_spec} and Fig.~S3). Subsequently, the fit of the resulting HWHM for the narrower Lorentzian $\gamma_{1}(q)$ with the jump diffusion model,~\cite{teixeira1985experimental} 
\begin{equation}
    \gamma_{1}(q)=\frac{D_s q^2}{1+\tau_{\mathrm{res}} D_s q^2}
    \label{jump}
\end{equation}
(see Figs.~S4, S5, S6) yields the self-diffusion coefficients $D_s$ and residence times $\tau_{\mathrm{res}}$ displayed as a function of ethanol mole fraction $x_{EtOD}$ in Fig.~\ref{fig_DandTau}. Herein values found on H$_{2}$O/C$_{2}$H$_{5}$OH mixtures in the absence of polarization analysis are reported for comparison.~\cite{seydel2019picosecond}
We find $D_s$ for pure H$_{2}$O and D$_{2}$O consistent with previously published values.~\cite{arbe2020coherent,teixeira1985experimental,mills1973self} 
The overall trend of $D_s$ for the partially deuterated mixture shows a minimum near 0.2 mole fraction ethanol corresponding to the maximum of the macroscopic viscosity and aligns with the results for the ensemble average self diffusion of fully protiated mixtures.~\cite{seydel2019picosecond}
Compared to those results, the larger decrease of the $D_s$ found in the present work is in good agreement with PFG-NMR results,~\cite{price1999self} where it is attributed to the H-bonded caging effect of water molecules around the alkyl group of ethanol.
It is remarkable that PA with selective deuteration allows to observe a similar trend for the self-diffusion of the ethyl group protons to that observed by NMR but with a higher time resolution (from \textit{ms} to \textit{ps}).
\begin{figure}[t!]
    \centering
        \centering
        {\includegraphics[width=\columnwidth]{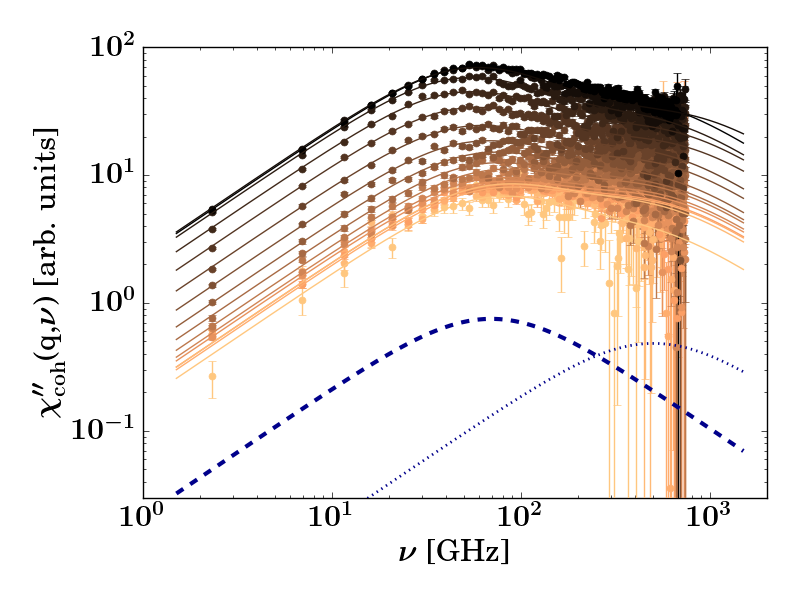}}
        \begin{picture}(0,0)
        \put(-50,150){(a)}        
        \end{picture}
        \label{relax}
    \hfill    
        \centering
        {\includegraphics[width=\columnwidth]{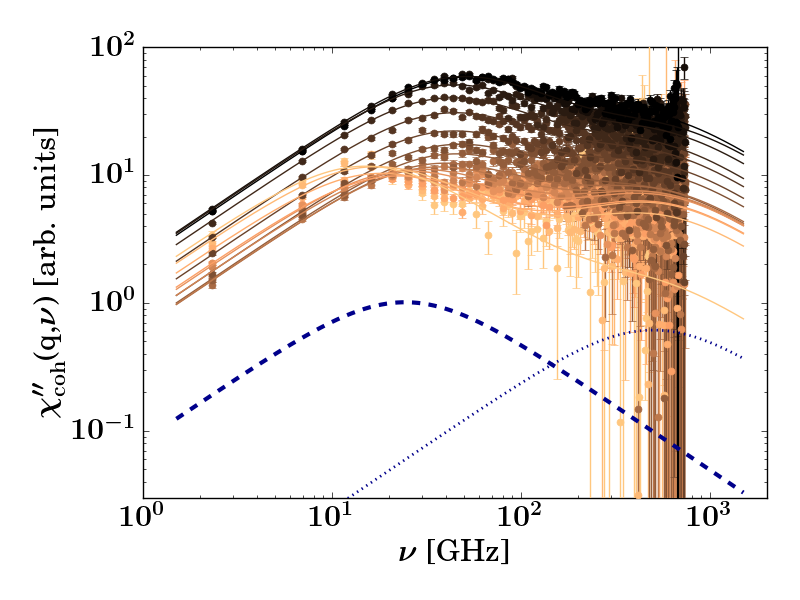}}
        \begin{picture}(0,0)
        \put(-50,150){(b)}        
        \end{picture}
        \label{amplitude}
    \caption{\footnotesize  Imaginary part of the dynamical susceptibility for the coherent spectrum of D$_{2}$O (a) and C$_2$H$_5$OD in D$_2$O at 0.02 ethanol mole fraction (b). Spectra recorded at 285K with E$_{i}$=3.84 meV in the range (orange) $0.4\,$\AA$^{-1}$$\leq q\leq 1.9\,$\AA$^{-1}$ (black). The ethanol rich sample shows a $q$-dependent peak at $\nu\sim10\,$GHz while for pure D$_{2}$O the maximum is almost $q$-independent. The solid lines stand for the fit according to eq.\ref{eq.Debye} with blue dashed and dotted lines depicting the main and the second Debye process (scaled by a factor 0.6 for visibility), respectively.}
    \label{fig_xi_fit_coh}
\end{figure}
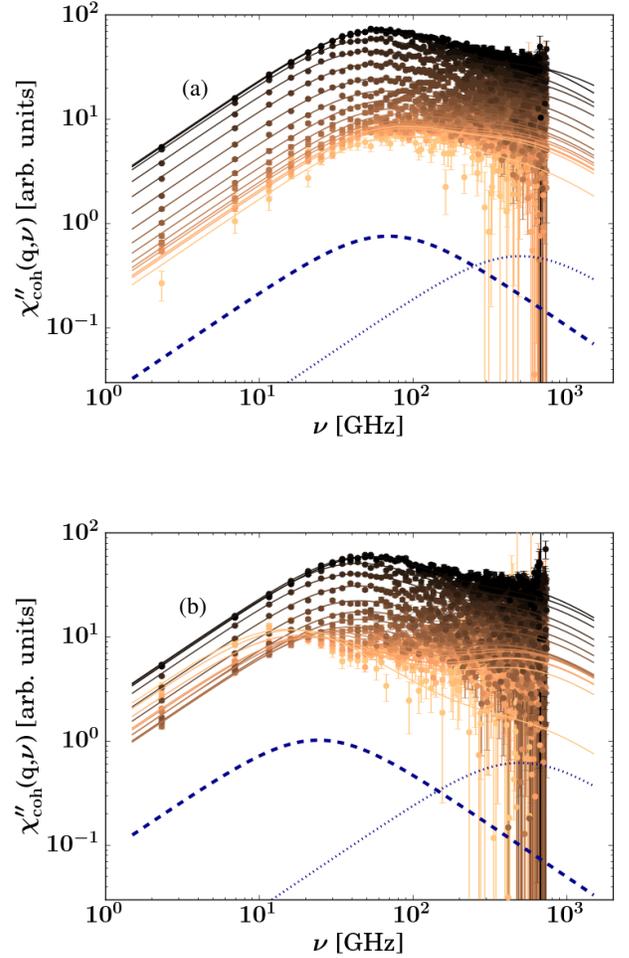  
\begin{figure}[t!]
    \centering
        \centering        {\includegraphics[width=\columnwidth]{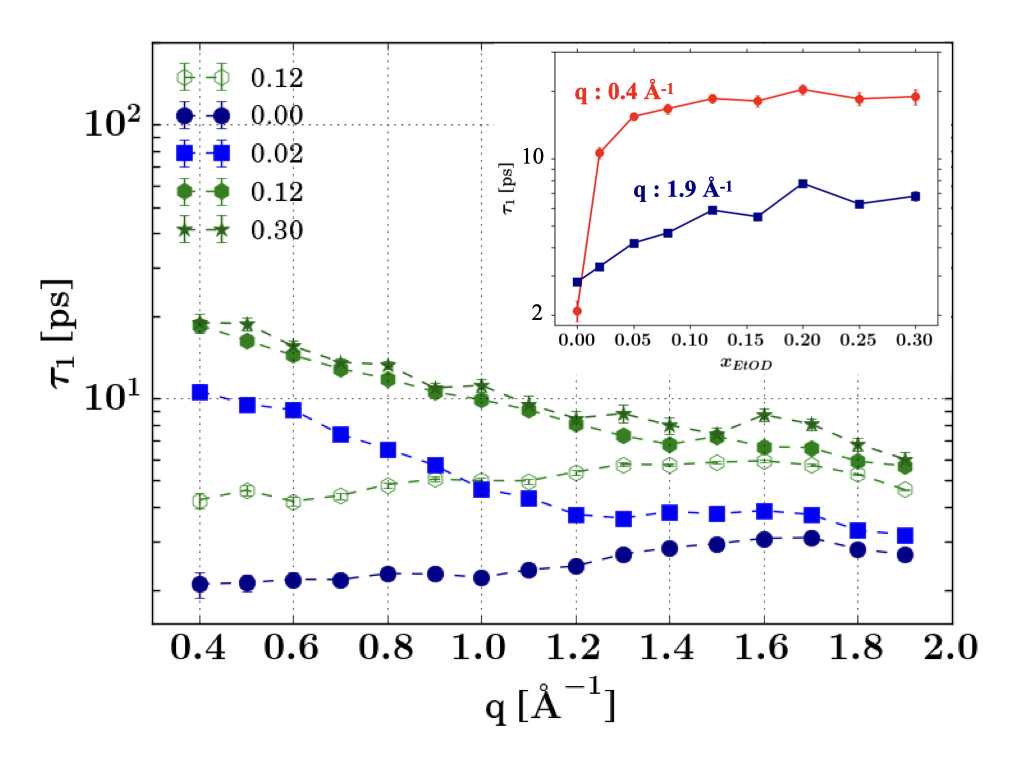}}
        \begin{picture}(0,0)
        \put(-30,150){(a)}        
        \end{picture}
    \hfill    
        \centering
        {\includegraphics[width=\columnwidth]{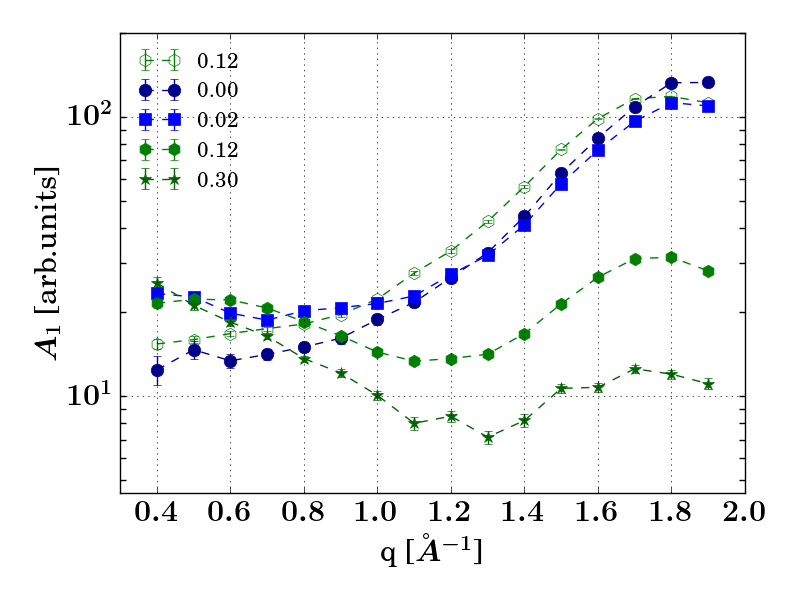}}
        \begin{picture}(0,0)
        \put(-30,150){(b)}        
        \end{picture}
    \caption{\footnotesize (a) Relaxation times $\tau_1(q)$ and (b) amplitude $A_1(q)$ of the main relaxation process of $\chi_{\mathrm{coh}}^{''}(q,\nu)$ as a function of ethanol mole fraction at T=285 K and for E$_1$=3.84 meV. Values obtained from the fit through eq.\ref{eq.Debye} and $\tau_1$ retrieved from the peak position $\nu_{i}^{\mathrm{max}}=(2\pi\tau_1)^{-1}$. \textit{Open symbols}: C$_2$D$_5$OD in D$_2$O, 0.12 ethanol mole fraction and T=290 K. \textit{Inset}: $\tau_1$ vs ethanol mole fraction at $q$=0.4 \AA$^{-1}$ (red) and $q$=1.9 \AA$^{-1}$ (blue).}
    \label{fitresult}
\end{figure}
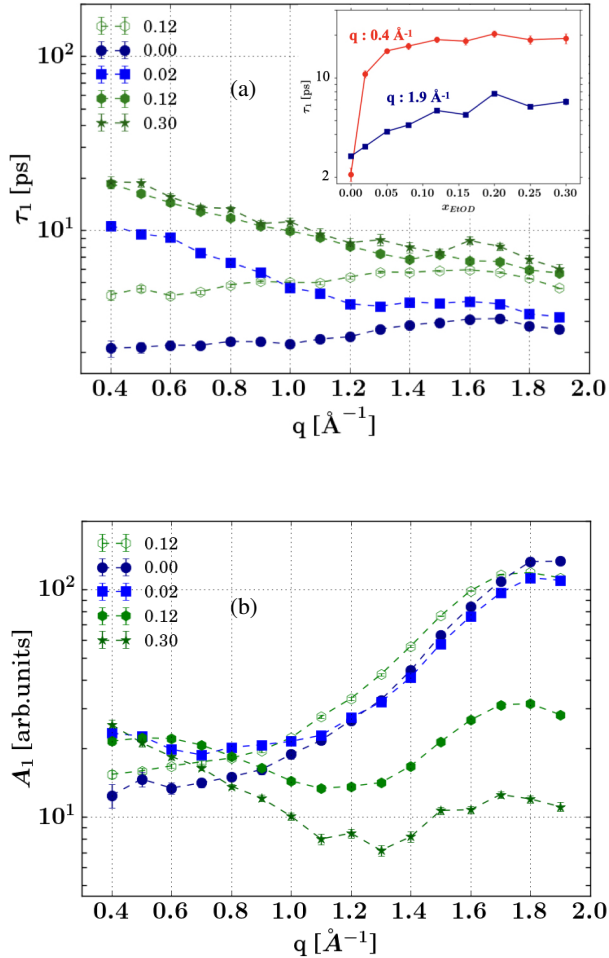 
We further analyse the coherent neutron scattering data in terms of the imaginary part of the dynamic susceptibility $\chi^{\prime\prime}(q,\nu)=\pi S(q,\nu)[1+n(\nu)]^{-1}$, where $\nu$ is the frequency and $n(\nu)$ the Bose occupation number given by $n(\nu)=(e^{h\nu/k_{B}T}-1)^{-1}$, allowing us to analyse the data in a model-free manner in the absence of analytical models describing the broadening of the coherent dynamic structure factor, especially at the mesoscale 
(cf.~Figs.~S7,~S8 and~S9).~\cite{novikov2013coherent,arbe2020coherent,bertrand2017nanoscopic,faraone2012coherent,burankova2014collective,Skold1967} Moreover, the conversion to dynamic susceptibility enhances the dynamic components of the spectrum displaying them as separated peaks. 
In the case of D$_{2}$O, $\chi_{\mathrm{coh}}^{\prime\prime}(q,\nu)$ displays two main dynamic features: a structural relaxation in the low frequency region 
($\nu<10^2\,$GHz) associated with non-vibrational motion and a higher frequency process linked to vibrations.~\cite{arbe2016dielectric,arbe2020coherent} 
Increasing the fraction of ethanol C$_2$H$_5$OD in D$_2$O, the first clear effect of perturbation of the pure water system brought on by ethanol is the appearance of the a peak in the low frequency region ($\nu\approx 10\,$GHz, Fig.~\ref{fig_xi_fit_coh}). To quantitatively model $\chi_{\mathrm{coh}}^{\prime\prime}(q,\nu)$ we use a linear combination of Debye-like relaxations,
\begin{equation}
    \chi_{\mathrm{coh}}^{\prime\prime}   \bigl(q,\nu\bigr)= \sum_{i} A_{i}(q)\frac{\nu\cdot\nu_{i}^{\mathrm{max}}(q)}{\nu^{2}+\nu_{i}^{\mathrm{max}}(q)^{2}}.
    \label{eq.Debye}
\end{equation} 
\textit{i.e.} Lorentzians in susceptibility with peak position $\nu_{i}^{\mathrm{max}}$ and
amplitudes $A_{i}(q)$ with index $i$, corresponding to exponential decays in the time domain with the relaxation time $\tau_{i}(q)=\big(2\pi \nu_{i}^{\mathrm{max}}(q)\big)^{-1}$.
In this picture we find two processes, as for pure D$_2$O: a main slow relaxation at low frequencies, also referred to as “$\alpha$-relaxation", and a second mode in the THz region (Fig.~\ref{fig_xi_fit_coh}). The resulting $\tau_{1}(q)$, associated with the peak position of the main relaxation process, is depicted in Fig.~\ref{fitresult} for the coherent part. Different models have been tested, with a variable number of generalized Debye processes (\textit{i.e.} Cole-Cole or Cole-Davidson relaxations) but the presented model with a sum of two Lorentzians provides the best match without over-fitting the experimental data.
For pure D$_{2}$O, we observe a $q$-independent relaxation $1\,$ps $\leq\tau_1\leq 3\,$ps up to 1.2 \AA$^{-1}$ followed by a maximum at $\sim$ 1.7 \AA$^{-1}$ associated with the D$_{2}$O nearest neighbour distance. This result shows a good agreement with the recent work on pure D$_2$O by Arbe \textit{et al.} who interpret the $q$-independent $\tau$ as consequence of the combination of terms making up the $S_{\mathrm{coh}}(q,\omega)$.~\cite{arbe2020coherent} This comprises three main components,
\begin{align}
    S_{\mathrm{coh}}(q,\omega) &=  \sum_{\alpha,\beta}b_{\rm coh}^{\alpha}b_{\rm coh}^{\beta} S_{\alpha\beta}(q,\omega) \\
    S_{\alpha\beta}(q,\omega) &= S^{\mathrm{self}}_{\alpha\alpha}(q,\omega)+S^{\mathrm{dist}}_{\alpha\alpha}(q,\omega)+S^{\alpha\neq\beta}_{\alpha\beta}(q,\omega)
    \label{selfdist}
\end{align}
where $S_{\alpha\beta}(q,\omega)$ are the partial dynamic structure factors of all pairs of atomic species $\alpha$ and $\beta$ (H, D, O and C for water-ethanol mixture) weighted by their coherent scattering lengths $b_{\rm coh}^{\alpha}$ and $b_{\rm coh}^{\beta}$; $S^{\mathrm{self}}_{\alpha\alpha}$ and $S^{\mathrm{dist}}_{\alpha\alpha}$ refers to the same and distinct particles of the same species $\alpha$, respectively. With the help of MD, Arbe \textit{et al.} interpreted the $q$-independent $\tau$ $\leq 1\, $\AA$^{-1}$ as a consequence of the cancelling out of $S^{\mathrm{dist}}_{\alpha\alpha}$ and $S^{\mathrm{self}}_{\alpha\alpha}$.~\cite{arbe2022understanding} In our work, we find that in the range between $0.02$ and $0.3$ ethanol mole fraction the relaxation times increase by up to one order of magnitude to $8\,$ps $\leq\tau_1\leq 20\,$ps (Fig.~\ref{fitresult}). This finding aligns to what has been shown for the same system by dielectric spectroscopy (DS) in the microwave region: the analysis of the dielectric loss for water-ethanol mixtures in terms of Debye relaxations reveals that $\tau_1$, associated with the structural reorganization of the HB network, ~\cite{cardona2018molecular,bohmer2014structure} linearly increases with ethanol mole fraction. According to the "wait-and-switch" model, ~\cite{cardona2018molecular,bohmer2014structure} the fluctuation of the hydrogen bond donors and acceptors between water and the -OH group are hindered by the reduction of available hydrogen bonding sites with increasing ethanol fraction. Our model fit for $\chi_{\mathrm{coh}}^{\prime\prime}(q,\nu)$ confirms this dynamical observation while adding structural information encoded in $q$.
\\
Two distinct regions can be distinguished in figure ~\ref{fitresult} also at lower incident energies (Fig.~S10): for $0.4\,$\AA$^{-1}$$\leq q\leq 1.2\,$\AA$^{-1}$ $\chi_{\mathrm{coh}}^{\prime\prime}(q,\nu)$ has a marked $q$ dependence while the slope observed for pure water in $1.2\,$\AA$^{-1}$$\leq q\leq 1.9\,$\AA$^{-1}$ flattens. The overall increase in the relaxation time is more pronounced in the low $q$ region and is not linearly related to mole fraction. Looking at $\tau_1$ as a function of ethanol mole fraction for two $q$ values it is clear how two different functional forms describe the relaxation time for meso- to intra-molecular scale (Fig.\ref{fitresult}a (inset)). We note that the susceptibility representation can lead to instabilities in the total spectral area calculation for some samples due to noise in the high frequency region (cf.~Fig.~S11). Importantly, the observed time scales may be influenced by limitations imposed by the spectrometer energy resolution (cf.~Fig.~S12) and
be reflecting maxima of distributions of times, as possibly expected for clusters
with a distribution of lifetimes.
\\
Interestingly, the $q$ range (0.4-1.2 \AA$^{-1}$) with a steeper increase for $\tau_1$ corresponds to the length scales where angular reorientation of the molecular dipoles probed by DS can be compared to collective density fluctuations measured by coherent QENS for water and its mixture.~\cite{arbe2018book} 
These density fluctuations may be associated with short-lived concentration fluctuations due to transient mesoscale inhomogeneities~\cite{subramanian2013mesoscale} seen, e.g., by simulations~\cite{ghosh2016temperature}, small-angle scattering~\cite{nishikawa1993small}, and indirectly inferred by thermodynamic analysis~\cite{takaizumi1997freezing}.
The distinct $q$-dependence of $\chi_{\mathrm{inc}}^{\prime\prime}(q,\nu)$ illustrates the key information from polarization analysis to access collective motions. Here, the relaxation time $\tau_{1}^{\mathrm{inc}}$ displays a strong $q$-dependence over the entire $q$ range consistent with the underlying diffusive process with a self-diffusion coefficient given by $D_{\mathrm{s}}\sim (\tau_1^{\mathrm{inc}} q^{2})^{-1}$ (Fig.~S13). While $\tau_1^{\mathrm{inc}}$ and $\tau_1^{\mathrm{coh}}$ for D$_{2}$O are one order of magnitude apart in the low $q$ region(Fig.~\ref{fitresult}(a), Fig.~S13),~\cite{arbe2020coherent} self- and collective relaxations for the binary mixture attain the same time window already at small ethanol fraction. Therefore the $q$-dependence of $\tau_1^{\mathrm{coh}}$ imply a diffusive process that is collective in nature as it arises from the pair-correlation function embodied in the coherent part of the spectrum in $\chi_{\mathrm{coh}}^{\prime\prime}(q,\nu)$. As such, the ``wait-and-switch" between different hydrogen-bonding sites is driven by approaching molecules at the ps time scale and neutron data with PA can directly probe these motions. 
In contrast to the increase in $\tau_1$, we observe a divergent trend for the per-deuteration (Fig.\ref{fitresult} open symbols): the relaxation profile D$_2$O/C$_2$D$_5$OD is closer to that of pure D$_{2}$O than to its partially deuterated D$_2$O/C$_2$H$_5$OD counterpart at the same ethanol mole fraction. This effect of deuteration might be linked to the interplay between the terms that make up the $S_{\mathrm{coh}}(q,\omega)$ (eq.\ref{selfdist}). It is likely that in the fully deuterated mixture the self and distinct terms cancel out similarly to the case of water,~\cite{arbe2020coherent} while for D$_2$O/C$_2$H$_5$OD they do not. This cancellation leads to the absence of the prepeak, expected in the mesoscale, in S$(q)$ which is obtained by integration over the final energy,
\begin{equation}
    S(q) = \int_{-\infty}^{\infty}S_{\mathrm{coh}}(q,\omega)d\omega
\end{equation}
(Fig.~S14). The prepeak usually appears as a shoulder on the low-$q$ side of the S$(q)$ maximum and, although its origin is still debated, in case of other lower alcohols is thought to be related to H-bonded molecular associates at distances longer than the first coordination shell.~\cite{ghoufi2020molecular,lenton2018temperature} There is an increasing consensus regarding the significance of deuteration in observing the prepeak.~\cite{chen2019wideangle} It has been shown that the prepeak in the coherent structure factor of methanol is only visible in CH$_3$OD and not in CD$_3$OD, ~\cite{bertrand2016dynamic,bertrand2017nanoscopic,chen2019wideangle} as has been shown by MD simulations to be based on the cancellation of the different partial factors for H and D depending on the $q$ range. The connection between structure and dynamics embodied in $\tau_1$ is clearly depicted in the plots of the amplitude $A_{1}(q)$, \textit{i.e.} the second free fit parameter of eq.\ref{eq.Debye}, for $\chi_{\mathrm{inc}}^{\prime\prime}(q,\nu)$ (Fig.~S13) and $\chi_{\mathrm{coh}}^{\prime\prime}(q,\nu)$ (Fig.~\ref{fitresult}b).  
While for the incoherent spectra $A_{1}(q)$ is constant in $q$ with only a non monotonous dependence on the ethanol mole fraction (see Fig.~S13), the amplitude of the main coherent relaxation process has an additional $q$-dependence (Fig.\ref{fitresult}b). In particular for D$_{2}$O, $A_{1}(q)$ matches the $S(q)$ shape~\cite{arbe2020coherent} with a gradual drop for $q\geq$ 1.1 \AA$^{-1}$ and conversely an increase for $q\leq$ 1.1 \AA$^{-1}$ upon ethanol addition. Even for $A_{1}(q)$ the fully deuterated mixture (open symbols) better aligns to the case of pure D$_{2}$O than to its partially deuterated counterpart. These findings for $A_{1}(q)$ are consistent with the picture of $S_{\mathrm{coh}}(q,\omega)$ being $S_{\mathrm{inc}}(q,\omega)$ modulated by the structure factor $S(q)$. 
\\ \\
In conclusion, employing the newly established sub-meV resolution polarization analysis and partial deuteration, we simultaneously measure: (i) self-diffusion of ethanol in the water matrix circumventing de Gennes narrowing; (ii) collective density fluctuations of the hydrogen bond network in the picosecond regime on a $q$ range from the meso- to intra-molecular scale. Compared to pure water, we observe a slower collective relaxation time upon increasing ethanol fraction with a steeper decrease in the mesoscale (0.4-1.2 \AA$^{-1}$) where the prepeak for lower alcohols has been observed. Coherent QENS reveals that the addition of small amounts of ethanol renders the hydrogen bond network significantly more rigid compared to pure water, consistent with long lived H-bonded cluster-like mesostructures.
By studying a binary molecular mixture, this work fills a gap around the H-bonded
fluids in the ps and mesoscale, 
that were previously only investigated as mono-molecular systems by high-resolution polarized QENS,\cite{arbe2020coherent,arbe2023collective} and provides
a non-MD dependent experimental basis suggesting the existence of clusters in water-ethanol mixtures.
\section*{Author Declarations}
\noindent {\bf Corresponding authors}\\
Tilo Seydel;\\ https://orcid.org/0000-0001-9630-1630;\\
Email: seydel@ill.eu\\
Katharina Edkins\\ https://orcid.org/0000-0002-6885-5457;\\
Email: katharina.edkins@manchester.ac.uk\\
\noindent {\bf Authors}\\
Riccardo Morbidini\\ https://orcid.org/0000-0003-4745-1610\\
Robert M.~Edkins\\ https://orcid.org/0000-0001-6117-5275 \\
Mark Devonport\\
G{\o}ran Nilsen\\ https://orcid.org/0000-0002-0148-0486
\subsection*{Author contributions}
All authors participated in the experiments (partially remotely due to COVID restrictions), data interpretation
and manuscript writing. RM analyzed the data with tutoring by TS. KE and TS designed and supervised the project.
\subsection*{Conflict of interest}
The authors have no conflicts to disclose.
\begin{acknowledgments}
This research has been supported by InnovaXN, a EU Horizon 2020 MSCA COFUND programme (\href{www.innovaxn.eu}{innovaxn.eu}, grant agreement No 847439). RM acknowledges an ILL PhD studentship in the ILL spectroscopy group funded by this program. We thank the STFC for allocation of beamtime on LET at the ISIS Neutron and Muon Facility, UK, and I.~Hoffmann (ILL) for a stimulating discussion on concentration fluctuations.
\end{acknowledgments}

\section*{Supplementary Material}
Introduction to neutron polarization analysis, spectrometer resolution, additional example spectra and fit results, apparent static structure factor from coherent dynamic scattering (PDF).

\section*{Data Availability Statement}
The neutron data are permanently curated by the ISIS facility and can be accessed by the experiment numbers RB1920548 and RB2210240, \url{https://doi.org/10.5286/ISIS.E.RB1920548} and \url{https://doi.org/10.5286/ISIS.E.RB2210240}.

\section*{References}
\bibliographystyle{aipnum4-1} 
\bibliography{references}
\end{document}